\documentclass[a4paper,11pt]{article}
\usepackage{pos}

\usepackage{color,xcolor}
\usepackage{xspace}

\newcommand{\be}{\begin{equation}}
\newcommand{\ee}{\end{equation}}
\newcommand{\bea}{\begin{eqnarray}}
\newcommand{\eea}{\end{eqnarray}}
\newcommand{\<}{\langle}
\renewcommand{\>}{\rangle}
\newcommand{\mc}{\mathcal}

\newcommand{\decay}[2]{\ensuremath{#1\!\to #2}\xspace}
\newcommand{\ellell}{\ensuremath{\ell^+\ell^-}}
\newcommand{\tautau}{\ensuremath{\tau^+\tau^-}}
\newcommand{\mumu}{\ensuremath{\mu^+\mu^-}}
\newcommand{\epen}{\ensuremath{e^+e^-}}
\newcommand{\epem}{\epen}
\newcommand{\nunu}{\ensuremath{\nu\bar{\nu}}}
\newcommand{\invfb}{\ensuremath{\mbox{\,fb}^{-1}\xspace}}
\newcommand{\Lb}{\ensuremath{\Lambda_b^0}}
\newcommand{\Bs}{\ensuremath{B_s^0}}
\newcommand{\Bd}{\ensuremath{B^0}}
\newcommand{\Bu}{\ensuremath{B^+}}
\newcommand{\Kp}{\ensuremath{K^+}}

\newcommand{\BF}{\ensuremath{{\cal B}}}
\newcommand{\Kstarz}{\ensuremath{K^{*0}}}
\newcommand{\Kstarp}{\ensuremath{K^{*+}}}

\newcommand{\vub}{\ensuremath{V_{\rm ub}}}

\newcommand{\vcb}{\ensuremath{V_{\rm cb}}}

\newcommand{\comment}[1]{}
\newcommand{\eg}{\textit{e.\,g.}}
\newcommand{\ie}{\textit{i.\,e.}}
\newcommand{\qsq}{\ensuremath{q^2}}
\newcommand{\deriv}{\ensuremath{\mathrm{d}}}
\newcommand{\tev}{\ensuremath{\mathrm{\,Te\kern -0.1em V}}\xspace}
\newcommand{\gev}{\ensuremath{\mathrm{\,Ge\kern -0.1em V}}\xspace}
\newcommand{\mev}{\ensuremath{\mathrm{\,Me\kern -0.1em V}}\xspace}
\newcommand{\kev}{\ensuremath{\mathrm{\,ke\kern -0.1em V}}\xspace}
\newcommand{\ev}{\ensuremath{\mathrm{\,e\kern -0.1em V}}\xspace}
\newcommand{\gevc}{\ensuremath{{\mathrm{\,Ge\kern -0.1em V\!/}c}}\xspace}
\newcommand{\mevc}{\ensuremath{{\mathrm{\,Me\kern -0.1em V\!/}c}}\xspace}
\newcommand{\gevcc}{\ensuremath{{\mathrm{\,Ge\kern -0.1em V\!/}c^2}}\xspace}
\newcommand{\gevgevcccc}{\ensuremath{{\mathrm{\,Ge\kern -0.1em V^2\!/}c^4}}\xspace}
\newcommand{\mevcc}{\ensuremath{{\mathrm{\,Me\kern -0.1em V\!/}c^2}}\xspace}

\title{WG3 Summary -- Rare $B$, $D$ and $K$ decays}

\author*[a]{Diego Guadagnoli}
\author[b]{Christoph Langenbruch}
\author[c]{Elisa Manoni}

\affiliation[a]{LAPTh, Universit\'{e} Savoie Mont-Blanc et CNRS, 74941 Annecy, France}

\affiliation[b]{RWTH Aachen University, I.\ Physikalisches Institut B, Sommerfeldstr.\ 14, 52056 Aachen, Germany}
\affiliation[c]{INFN Sezione di Perugia, I-06123 Perugia, Italy}

\emailAdd{diego.guadagnoli@lapth.cnrs.fr}
\emailAdd{christoph.langenbruch@cern.ch}
\emailAdd{elisa.manoni@pg.infn.it}

\abstract{
We summarize the presentations made within Working Group 3 of the CKM2021 workshop. This working group is devoted to rare $B$, $D$ and $K$ decays, radiative and electroweak-penguin decays, including constraints on $V_{\rm td}/V_{\rm ts}$ and $\epsilon^\prime / \epsilon$. The working group has thus a very broad scope, and includes very topical subjects such as the coherent array of discrepancies in semi-leptonic $B$ decays. Each contribution is here summarized very succinctly with the aim of providing an overview of the main results. The reader interested in fuller details is referred to the individual contributions.
}

\FullConference{%
  11${}^{th}$ International Workshop on the CKM Unitarity Triangle (CKM2021)\\
  22-26 November 2021\\
  The University of Melbourne, Australia
}

\begin{document}
\maketitle

\clearpage
\section{Introduction}

\noindent Rare $B$, $D$ and $K$ decays are so denoted because they are very suppressed within the Standard Model (SM), by various mechanisms including at least loop and CKM suppression. As a consequence, effects from new particles beyond the SM (New Physics, or NP for short) can result in significant deviations from SM predictions. These processes thereby constitute powerful probes of phenomena beyond the SM. 
The search for NP effects requires high precision from both experimental measurements and theory predictions, and this is often possible for rare meson decays.
In the following, we will summarise the state of the field as presented at the "$11^{th}$ International Workshop on the CKM unitarity triangle" (CKM 2021). 
For details, the reader is kindly referred to the individual contributions to these proceedings. 

\section{\boldmath Rare $B$ decays}
\subsection{Rare leptonic decays}
\noindent The purely leptonic decays $\decay{\Bs}{\mumu}$ and $\decay{\Bd}{\mumu}$ are loop-, helicity- and CKM-suppressed in the SM, and they are very sensitive to potential contributions from new scalar sectors. 
In the SM, the branching fractions of the decays are precisely predicted to be~\cite{Bobeth:2013uxa,Beneke:2019slt}
\begin{align}
    \BF(\decay{\Bs}{\mumu}) &= (3.66\pm 0.14)\times 10^{-9}~,\\
    \BF(\decay{\Bd}{\mumu}) &= (1.03\pm 0.05)\times 10^{-10}.\nonumber
\end{align}
One of the leading contributions to the uncertainty of the SM prediction originates from the uncertainty on the CKM matrix element \vcb. 
As presented by E.\ Venturini at this workshop~\cite{Buras:2022nrb}, it is possible to define clean ratios of observables in $B$ decays and mixings, in which parametric uncertainties largely cancel. 
An important example is the ratio 
\begin{align}
R_{s(d)\mu} &\equiv \frac{\BF(\decay{B_{s(d)}^0}{\mumu})}{\Delta M_{s(d)}}
\end{align}
in which the \vcb\ dependence cancels and which thereby can be predicted without this dependence~\cite{Buras:2003td}. Using this ratio, and an average of 2+1 and 2+1+1 lattice results (see Ref.~\cite{Aoki:2021kgd} for a complete discussion), Ref.~\cite{Bobeth:2021cxm} finds agreement with the SM at no better than 2.1$\sigma$, which is increased to $2.7\sigma$ if one uses 2+1+1 results only \cite{Buras:2022wpw}. 
One can systematically work out  many of such \vcb-independent ratios~\cite{Buras:2021nns}, and identify their leading CKM dependence, which may be parameterized \eg\ in terms of the angles $\beta$ and $\gamma$. Such an approach may allow for an assessment of the big picture that is less reliant on parametric errors that are still under debate. 

\begin{figure}
    \centering
    \includegraphics[height=0.3\textwidth]{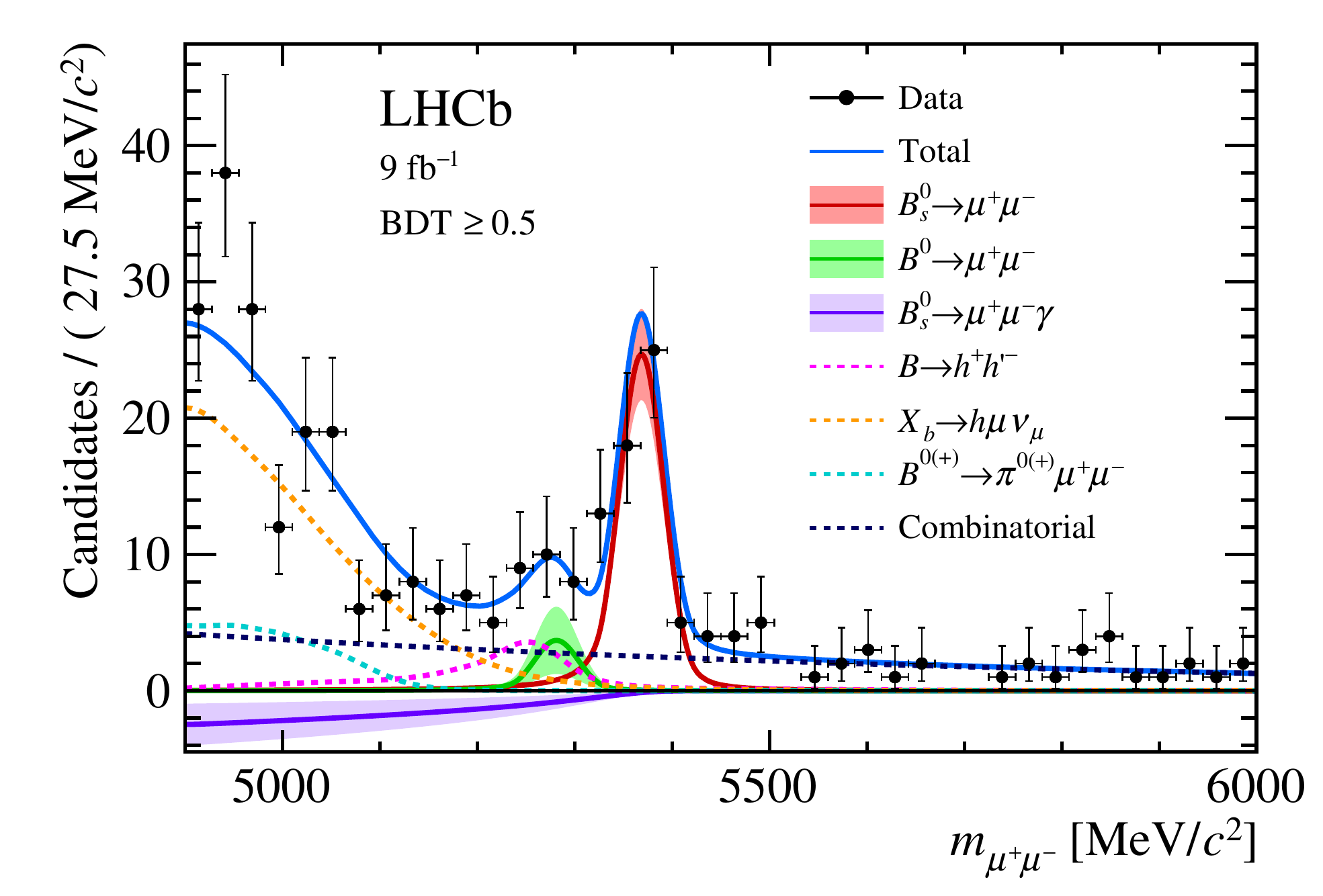}
    \includegraphics[height=0.3\textwidth]{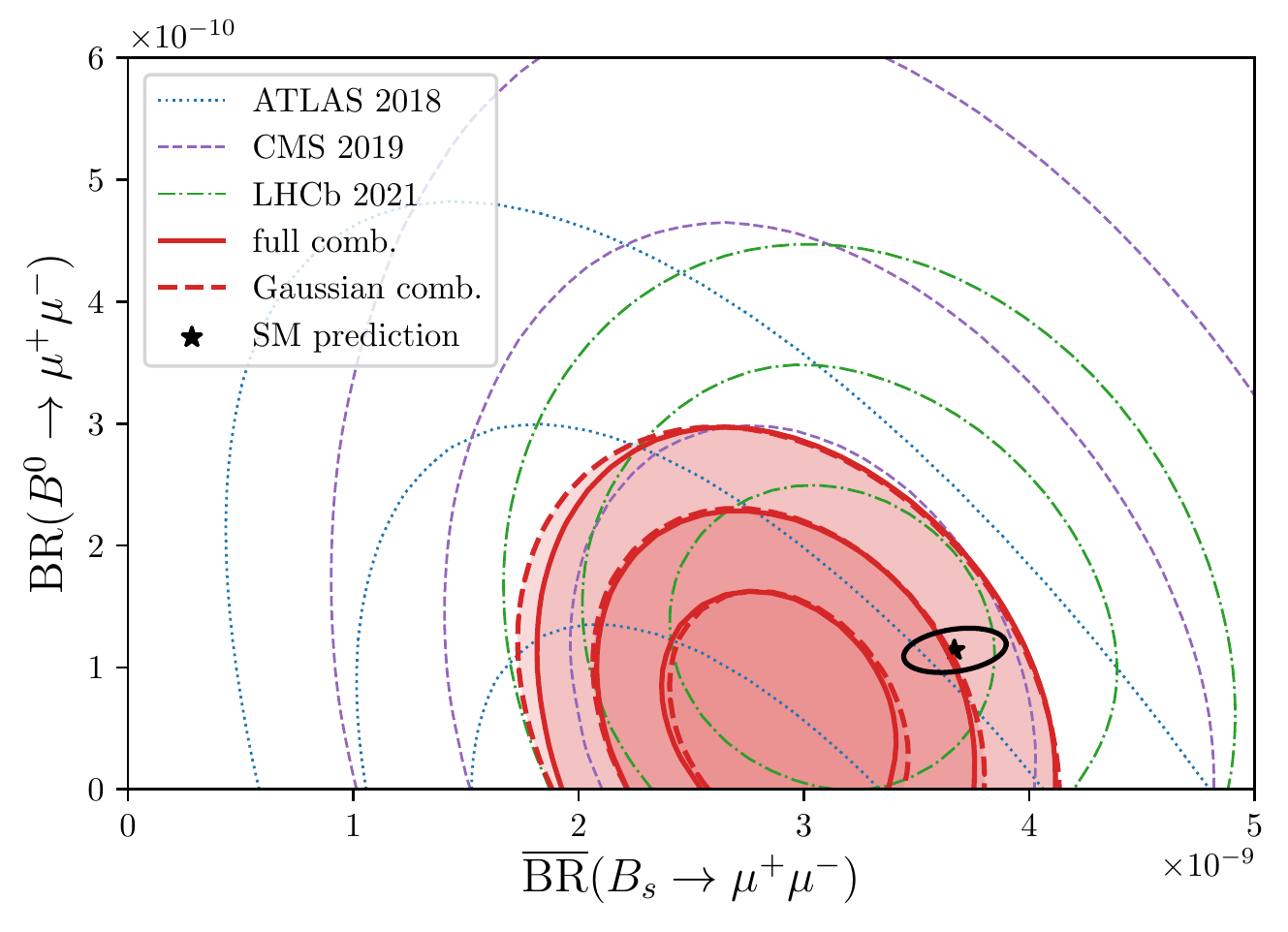}
    \caption{(Left) Invariant mass of the dimuon system for $\decay{B_{(s)}^0}{\mumu}$ signal candidates~\cite{LHCb:2021vsc}. (Right) Combined results of the data by LHCb, ATLAS and CMS on $\BF(\decay{\Bs}{\mumu})$ and $\BF(\decay{\Bd}{\mumu})$~\cite{Altmannshofer:2021qrr}.}
    \label{fig:bs2mumu}
\end{figure}

The latest measurements of $\BF(\decay{\Bs}{\mumu})$ and $\BF(\decay{\Bd}{\mumu})$ by LHCb were presented by F.\ Dettori~\cite{proceedings:dettori}. The analysis uses the full LHCb Run 1 and 2 data sample, the resulting distribution of the invariant mass of the dimuon system is shown in Fig.~\ref{fig:bs2mumu} (left). The measured branching fractions~\cite{LHCb:2021vsc,LHCb:2021awg} 
\begin{align}
    \BF(\decay{\Bs}{\mumu}) &= (3.09^{+0.46}_{-0.43}{}^{+0.15}_{-0.11})\times 10^{-9}~,\\
    \BF(\decay{\Bd}{\mumu}) &= (1.2^{+0.8}_{-0.7}\pm 0.1)\times 10^{-10},\nonumber
\end{align}
are in good agreement with the SM predictions. 
In addition, the first limit on the decay $\decay{\Bs}{\mumu\gamma}$ was obtained using a new method~\cite{Dettori:2016zff}. This first limit provides guidance to recent theoretical studies about its SM prediction and NP potential  \cite{Guadagnoli:2017quo,Kozachuk:2017mdk,Beneke:2020fot,Carvunis:2021jga}. 
In this context, it is also worth pointing out a new precise measurement of the hadronisation fraction ratio $f_s/f_d$~\cite{LHCb:2021qbv}, which can otherwise limit measurements of branching fractions of \Bs\ decays at LHCb (see Ref.~\cite{Bordone:2020gao} for a recent theory reappraisal of this subject).

The current status of measurements of rare decays of heavy flavour by ATLAS and CMS was presented by P.~Reznicek~\cite{proceedings:reznicek}.  
Both collaborations have performed measurements of $\BF(\decay{\Bs}{\mumu})$ and $\BF(\decay{\Bd}{\mumu})$~\cite{ATLAS:2018cur,CMS:2019bbr}, which have been combined with the LHCb measurement~\cite{LHCb:2021vsc,LHCb:2021awg} in Ref.~\cite{Altmannshofer:2021qrr}.  
The combination of the results is shown in Fig.~\ref{fig:bs2mumu} (right) and shows a tension with the SM prediction at the level of $2\,\sigma$. 

Searches for the lepton-flavour violating decays $\decay{\Bd}{\tau^\pm\mu^\mp}$ and $\decay{\Bd}{\tau^\pm e^\mp}$ have been performed at Belle, as presented by T.~Luo~\cite{proceedings:luo}. Upper limits of 
$\BF(\decay{\Bd}{\tau^\pm\mu^\mp})<1.5\times 10^{-5}$ and $\BF(\decay{\Bd}{\tau^\pm e^\mp})<1.6\times 10^{-5}$ have been set at 90\%~CL~\cite{Belle:2021rod}.

\subsection{Rare semileptonic decays} \label{sec:rare_semileptonic_decays}
\begin{figure}
    \centering
    \includegraphics[width=0.45\textwidth]{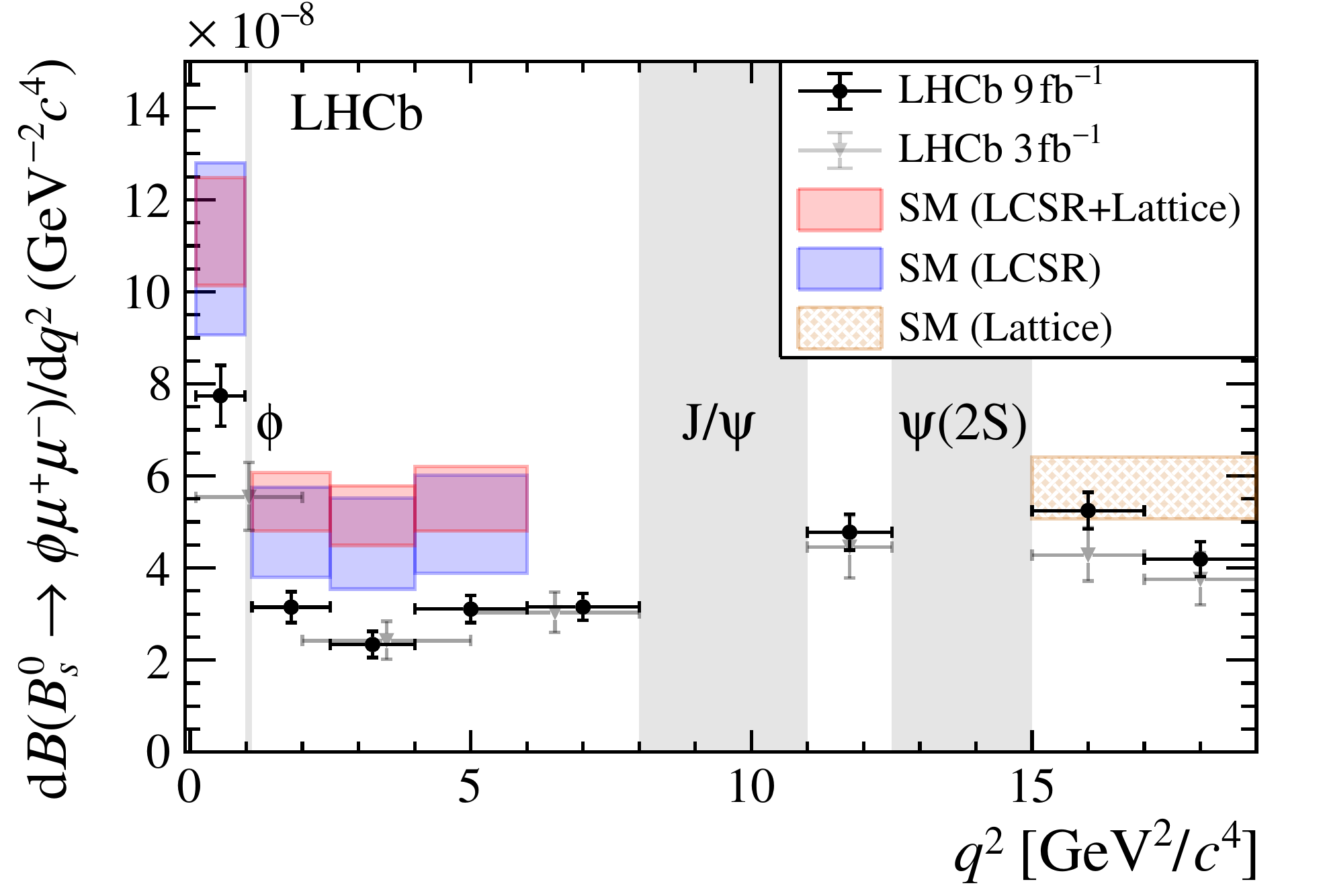}
    \includegraphics[width=0.45\textwidth]{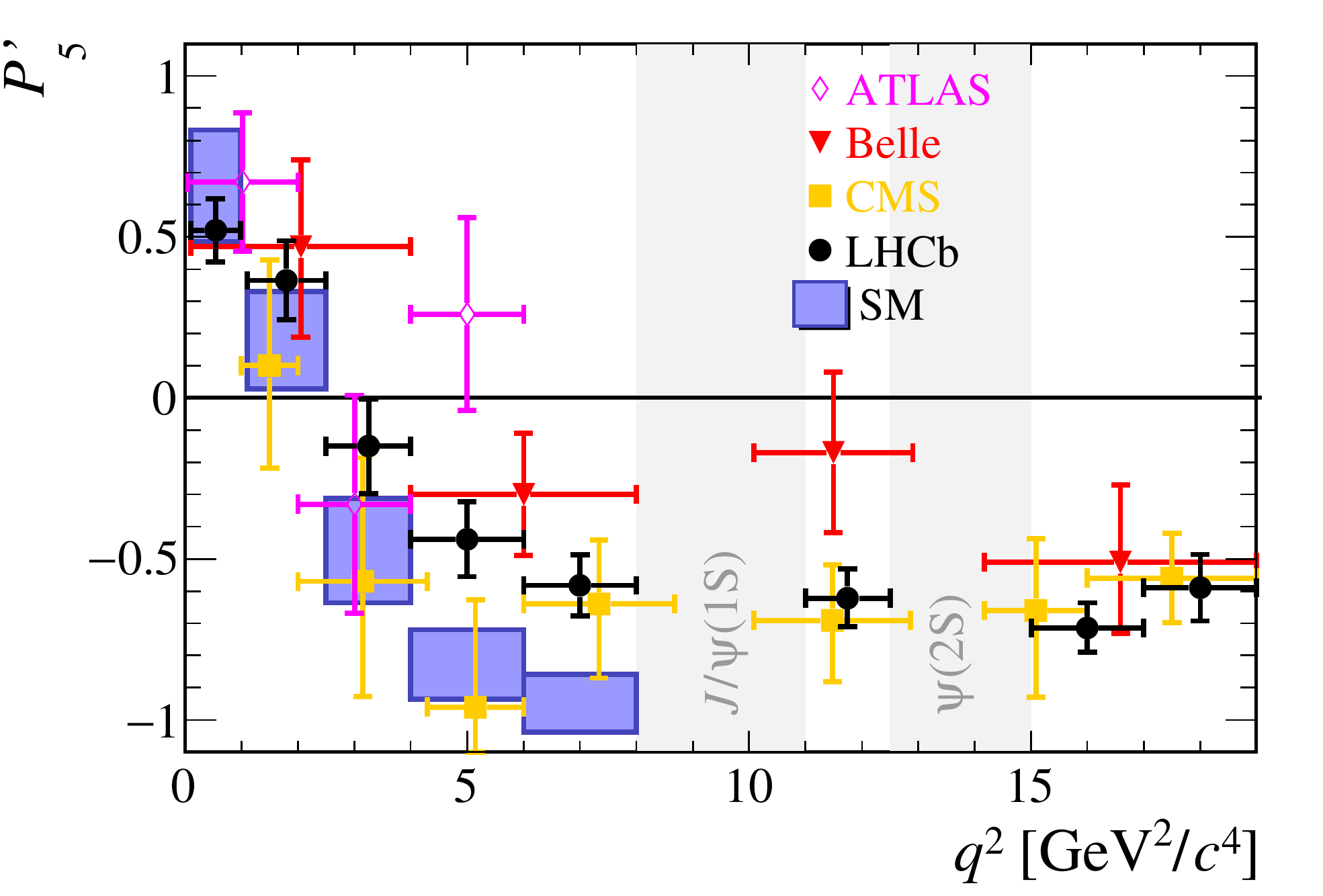}
    \caption{(Left) Differential branching fraction of the rare decay $\decay{\Bs}{\phi\mumu}$~\cite{LHCb:2021zwz}. (Right) Overview of results on the angular observable $P_5^\prime$ in the decay $\decay{\Bd}{\Kstarz\mumu}$~\cite{LHCb:2020lmf,Belle:2016fev,ATLAS:2018gqc,CMS:2017rzx}. Figure from Ref.~\cite{Albrecht:2021tul}.}
    \label{fig:b2smumu}
\end{figure}

\noindent Rare semileptonic $\decay{b}{s\mumu}$ decays are loop- and CKM-suppressed in the SM and thus NP can significantly affect the branching fractions of these processes and the angular distributions of the final state particles. 
M.~Kreps~\cite{proceedings:kreps} presented a review of the results on $\decay{b}{s\mumu}$ decays at LHCb. 
Several intriguing tensions exist in this area, which persist in the present. 
The most recent results by LHCb on $\decay{b}{s\mumu}$ decays have been obtained using the decay $\decay{\Bs}{\phi\mumu}$~\cite{LHCb:2021zwz,LHCb:2021xxq}. 
The branching fraction of this decay is found to lie $3.6\,\sigma$ below the SM prediction~\cite{LHCb:2021zwz}, as shown in Fig.~\ref{fig:b2smumu} (left). 
The angular distributions of the decay are found to be in tension with the SM prediction at the level of $2\,\sigma$, determined using a global analysis of the decay~\cite{LHCb:2021xxq}. 
These tensions are consistent with similar discrepancies observed in the angular analyses of the decays $\decay{\Bd}{\Kstarz\mumu}$~\cite{LHCb:2020lmf} and $\decay{\Bu}{\Kstarp\mumu}$~\cite{LHCb:2020gog}. 
A recent angular analysis of the decay $\decay{\Bu}{\Kstarp\mumu}$, using the full Run~1 and~2 data sample collected by LHCb, reveals a tension corresponding to $3.1\,\sigma$ with the SM prediction. 
This is consistent with an earlier angular analysis 
of $\decay{\Bd}{\Kstarz\mumu}$ which exhibits a $3.3\,\sigma$ tension~\cite{LHCb:2020lmf}. 
The angular observable $P_5^\prime$ determined by LHCb in this analysis is shown in Fig.~\ref{fig:bs2mumu} (right), together with results by the Belle, ATLAS, and CMS collaborations~\cite{Belle:2016fev,ATLAS:2018gqc,CMS:2017rzx}. 
In summary, this set of measurements suggests deviations going consistently in one and the same direction. Further measurements, and updates of existing ones, are eagerly anticipated. 

The significance of the tensions with SM predictions in semileptonic $\decay{b}{s\ellell}$ decays depends on the precision of the SM prediction. 
Danny van Dyk discussed recent progress on theory errors in this area~\cite{proceedings:vandyk}. The theory errors are dominated by the knowledge of the QCD dynamics underlying the required matrix elements. Starting from the general $|\Delta B| = |\Delta S| =  1$ effective Lagrangian, the most relevant subset of operators includes: the semileptonic operators, known as $\mathcal{O}_{9,10}$; the 4-quark operators with two charm quarks, known as $\mathcal{O}_{1,2}^c$; and the electromagnetic dipole operator $\mathcal{O}_7$. The speaker notes that, in order to identify the main long-distance quantity to calculate, one can write down a mock amplitude, consisting of terms that are the product of local form factors $\mc F_\lambda$ times the Wilson coefficients of the mentioned operators, plus terms proportional to non-local form factors $H_\lambda$ (where $\lambda$ denotes the helicity) of dimension-5 operators.
Starting from such an expansion, the crucial expression is the $T$-product
\be
\< H_s| \int d^4 x e^{i q \cdot x} \mathcal T \{ j^\mu_{\rm e.m.}(x), [C_1 \mathcal O_1^c + C_2 \mathcal O_2^c](0)\} | H_b \>~.
\ee
The state-of-the-art until recently was the well-known calculation by KMPW~\cite{Khodjamirian:2010vf}, where the authors expanded the $T$-product in light-cone operators under the assumption $q^2 - 4 m_c^2 \ll \Lambda_{\rm had} m_b$, then expressed the leading contribution through local form factors $\mc F_\lambda$. Within this approach, $1 / (q^2 - 4 m_c^2)$ corrections can, in principle, be systematically obtained.
The speaker and his collaborators \cite{Bobeth:2017vxj} put in place a new strategy, where one computes $H_\lambda$ at space-like $q^2$, and then extrapolate to timelike $q^2 \le 4 M_D^2$ by way of a suitable parameterization. Importantly, this strategy includes information from non-leptonic decays with $J/\psi$ and $\psi(2S)$ final states. This method informed the recent reappraisal of KMPW performed in Ref.~\cite{Gubernari:2020eft} (see also Ref.~\cite{Gubernari:2018wyi}). A full analysis is work-in-progress.

New physics entering $b\to s\mumu$ process can also affect $b\to s$ transition accompanied by tau and neutrino pairs.
Results on semileptonic decays from the Belle and Belle II collaborations were presented by T.~Luo~\cite{proceedings:luo}. 
Semileptonic $\decay{b}{s\tautau}$ decays are of particular interest as effects from NP can be large for the third generation leptons \cite{Glashow:2014iga,Capdevila:2017iqn}. 
The Belle collaboration recently performed a search for the decay $\decay{\Bd}{\Kstarz\tautau}$, using the full data sample~\cite{Belle:2021ndr}. 
No significant signal is observed and an upper limit of 
$\BF(\decay{\Bd}{\Kstarz\tautau}) < 2.0\times 10^{-3}$ 
at 90\% confidence level (CL) is set, which constitutes the first experimental constraint on this mode~\cite{Belle:2021ndr}. 
In addition, a search for the decay $\decay{\Bu}{\Kp\nunu}$ using $63\invfb$ of Belle~II data was presented~\cite{Belle-II:2021rof}. 
The analysis uses a novel inclusive tagging approach, 
inspired by Ref.~\cite{Belle:2019iji}, which exhibits larger signal efficiency than previous analyses with hadronic or semileptonic tag.  
The search measures a branching fraction of $\BF(\decay{\Bu}{\Kp\nunu})=1.9^{+1.6}_{-1.5}\times 10^{-5}$ and establishes an upper limit of $\BF(\decay{\Bu}{\Kp\nunu})<4.1\times 10^{-5}$ at 90\%~CL. 

\subsection{\boldmath Radiative $B$ decays}
\noindent Radiative $\decay{b}{s(d)\gamma}$ decays constitute flavour changing neutral currents that are highly sensitive to NP contributions. 
Both inclusive measurements of radiative $b$-hadron decays and measurements of branching fractions and CP asymmetries of exclusive final states are of great interest and give important complementary information.  

The theory status of inclusive $\bar B \to X_s \gamma$ decays was reviewed by A.~Rehman~\cite{proceedings:rehman}. 
The CP- and isospin-averaged branching fraction $\BF(\decay{B}{X_s\gamma})$ is known experimentally to $4.5\%$ precision~\cite{HFLAV:2019otj}. 
This experimental result is given for $E_\gamma > 1.6$ GeV as customary. Since experimental backgrounds increase for lower $E_\gamma$, measurements are obtained in the range [1.7, 2.0] GeV and then extrapolated to 1.6 GeV. On the theory side one wants a photon energy $E_0$ large ($\sim m_b / 2$), in order to keep $m_b - 2 E_0 \gg \Lambda_{\rm QCD}$. This justifies the choice of $E_0 = 1.6$ GeV. 
The theory calculation of the branching fraction consists of a perturbative (96\% of the prediction) and of a non-perturbative (4\%) component. The former component is the partonic width integrated over $E_\gamma > E_0$, and calculated as an expansion bilinear in Wilson coefficients $C_i(\mu_b) C_j(\mu_b)$, weighted by functions $\hat G_{i,j}(E_0,\mu_b)$. The $G_{27}^{(2)}$ term, \ie\ the interference between the operators known as $\mathcal O_2$ and $\mathcal O_7$, is considered to NNLO in QCD and for arbitrary charm mass~\cite{Misiak:2020vlo}. With respect to the previous state-of-the-art calculation~\cite{Misiak:2015xwa}, this implies a sizeable reduction of the error (from about 7\% to 5\%)  for both $\BF_{s\gamma}^{\rm SM}$ and for $R_\gamma \equiv {\BF}_{(s+d) \gamma} / {\BF}_{c \ell \bar \nu}$. The theory error is, however, still dominant with respect to the ultimate expected Belle II accuracy of about 3\% with $50\,\mathrm{ab}^{-1}$~\cite{Belle-II:2018jsg}.

Recent measurements of radiative $b$-hadron decays by LHCb were presented by C.~Mar\'in Benito~\cite{proceedings:marinbenito}. 
A time-dependent tagged analysis of the exclusive decays $\decay{\Bs}{\phi\gamma}$ gives access to the observables ${\cal A}^{\Delta}_{\phi\gamma}$, $S_{\phi\gamma}$, and $C_{\phi\gamma}$, which are sensitive to the photon polarisation. 
The resulting values of~\cite{LHCb:2019vks} 
\begin{align*}
    {\cal A}^{\Delta}_{\phi\gamma} &= -0.67^{+0.37}_{-0.41}\pm0.17,\\
S_{\phi\gamma} &= 0.43\pm 0.30\pm 0.11~,\\
C_{\phi\gamma} &= 0.11\pm 0.29\pm 0.11~,
\end{align*}
are in good agreement with the SM prediction and a previous untagged measurement~\cite{LHCb:2016oeh}. 
LHCb also allows to study exclusive radiative $b$-baryon decays. 
The first observation of a radiative $b$-baryon decay was performed by LHCb with the observation of the decay $\decay{\Lb}{\Lambda\gamma}$~\cite{LHCb:2019wwi}. 
An angular analysis of this mode~\cite{Mannel:1997pc,Hiller:2001zj} allowed recently to measure the photon polarization  $\alpha_\gamma=0.82^{+0.17}_{-0.26}{}^{+0.04}_{-0.13}$~\cite{LHCb:2021byf}, which is in good agreement with the SM prediction. 
A search for the radiative $b$-baryon decay $\decay{\Xi_b^-}{\Xi^-\gamma}$ results in an upper limit of $\BF(\decay{\Xi_b^-}{\Xi^-\gamma})<1.3\times 10^{-4}$ at 95\% CL~\cite{LHCb:2021hfz}. 

Recent measurements and prospects on the $b\to s \gamma$ sector at the Belle and Belle II experiments were presented by M.~Röhrken~\cite{proceedings:roehrken}.
Belle analyzed the $771\invfb$ full dataset providing the most stringent measurements on branching fraction and CP asymmetries for the $B \to K^{\ast} \gamma$ channel, along with a measurement of the isospin asymmetry showing a $3.1~\sigma$ evidence for isospin violation~\cite{Belle:2017hum}. Belle II performed a search for the same channel on about $63~\invfb$ providing branching fraction measurements~\cite{BelleII:2021tzi}. With larger statistics, updated estimations of the rates along with the isospin and CP asymmetries will be performed. With a sample of $5\,\mathrm{ab}^{-1}$ accuracies at sub-percent level are expected on all aforementioned quantities. Time-dependent analyses to determine the photon polarization will also be feasible~\cite{Belle-II:2018jsg}.     
On the inclusive side, several techniques to reconstruct $B \to X_s \gamma$ final states, boiling down to different assumptions on the tag-side decay, are feasible at $B$ factories. The different assumptions on the tag side (from no assumption for a fully inclusive analysis to the request of a fully reconstructed hadronic $B$ mode) translate into a trade-off between efficiencies and purities. 
Belle~II presented a measurement of the photon spectrum with fully inclusive tag method and no requirement on the $X$ system accompanying the high energy cluster, observing an excess in background-subtracted data in the signal-like region. This represents a starting point for the wide $B \to X \gamma$ program Belle~II will carry out~\cite{Belle-II:2018jsg}.

\subsection{Tests of lepton universality in rare decays}
\begin{figure}
    \centering
    \includegraphics[height=0.3\textwidth]{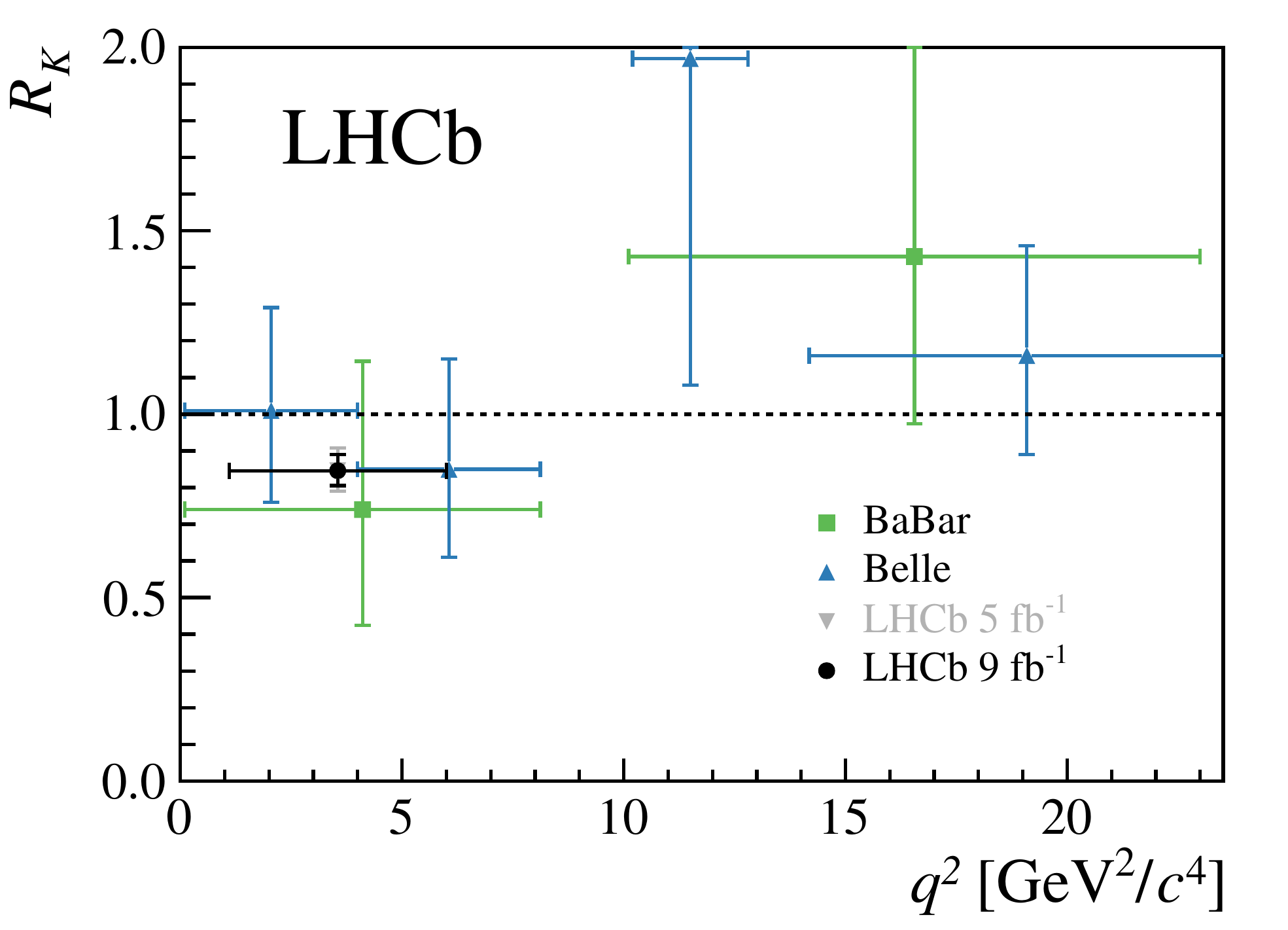}
    \includegraphics[height=0.3\textwidth]{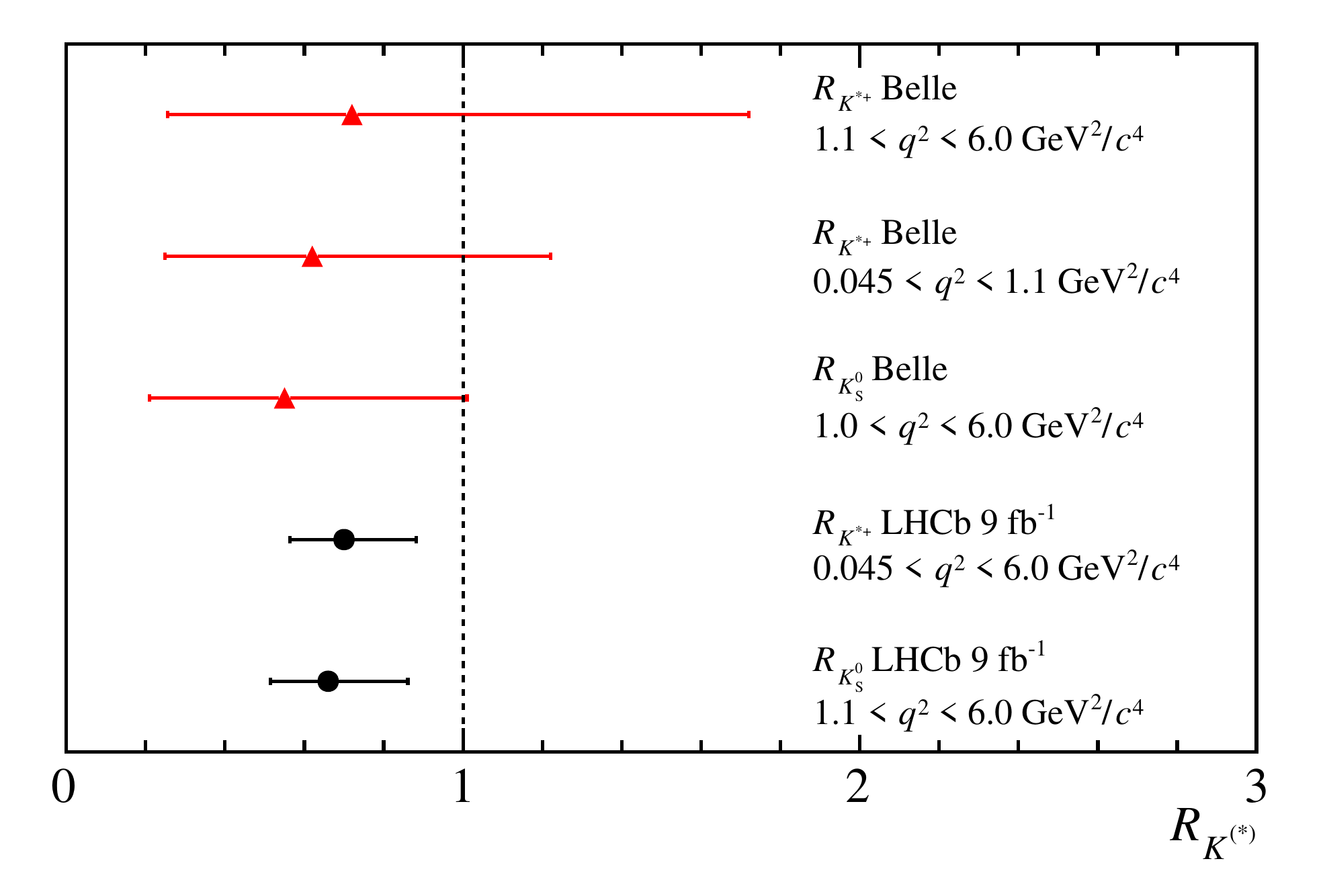}
    \caption{(Left) Measurements of $R_K$ by LHCb~\cite{LHCb:2021trn,LHCb:2019hip}, BaBar~\cite{BaBar:2012mrf}, and Belle~\cite{Belle:2009zue}. (Right) Lepton flavour universality tests $R_{K^{*+}}$ and $R_{K_S^0}$ by LHCb~\cite{LHCb:2021lvy} and Belle~\cite{Belle:2019oag,BELLE:2019xld}. Figures from Refs.~\cite{LHCb:2021trn,LHCb:2021lvy}.}
    \label{fig:lfuexp}
\end{figure}

\noindent The SM interactions are to an excellent approximation lepton-flavour universal. 
Lepton flavour universality (LFU) can be tested using the ratios
\begin{align}
    R_H &= \frac{\int_{\qsq_\mathrm{min}}^{\qsq_\mathrm{max}} \frac{\deriv\BF(\decay{B}{H\mumu})}{\deriv\qsq}\deriv\qsq}{\int_{\qsq_\mathrm{min}}^{\qsq_\mathrm{max}} \frac{\deriv\BF(\decay{B}{H\epem})}{\deriv\qsq}\deriv\qsq},
\end{align}
which are precisely predicted to be unity in the SM if lepton masses can be neglected. 
QED corrections are at most of ${\cal O}(1\%)$~\cite{Bordone:2016gaq}, and, crucially, hadronic effects cancel in the ratio. Any significant departure from unity would therefore provide a clear hint of NP.  
The most precise measurement of $R_K$ is performed by the LHCb collaboration using the full Run~1 and~2 data sample~\cite{LHCb:2021trn}. 
The resulting value of
\begin{align}
    R_K(1.1<\qsq<6\gevgevcccc) &= 0.846_{-0.039}^{+0.042}{}_{-0.012}^{+0.013},
\end{align}
shown in Fig.~\ref{fig:lfuexp} (left), 
is below the SM prediction, the tension corresponds to $3.1\,\sigma$~\cite{LHCb:2021trn}. 
Measurements by the BaBar and Belle collaborations exhibit larger uncertainties and are compatible with both the SM and the LHCb measurement~\cite{BaBar:2012mrf,Belle:2009zue}. 
More recently, the LHCb collaboration performed a measurement of the LFU tests $R_{K_S^0}$ and $R_{K^{*+}}$~\cite{LHCb:2021lvy}, the resulting values of
\begin{align}
    R_{K_S^0}(1.1<\qsq<6\gevgevcccc) &= 0.66^{+0.20}_{-0.15}{}^{+0.02}_{-0.04}\\
    R_{K^{*+}}(0.045<\qsq<6\gevgevcccc) &= 0.70^{+0.18}_{-0.13}{}^{+0.03}_{-0.04}\nonumber
\end{align}
are shown in Fig.~\ref{fig:lfuexp} (right), they lie below the SM prediction at $1.5\,\sigma$ and $1.4\,\sigma$, respectively~\cite{LHCb:2021lvy}. 
Measurements by the Belle collaboration are consistent with either the SM and LHCb~\cite{Belle:2019oag,BELLE:2019xld}. 
Earlier measurements of $R_{K^*}$ and $R_{pK}$ by the LHCb collaboration were also found to lie below the SM prediction~\cite{LHCb:2017avl,LHCb:2019efc}. 
The pattern of measurements of clean lepton flavour universality ratios below the SM prediction is intriguingly consistent with that observed in branching ratios and in the $\decay{\Bd}{\Kstarz\mumu}$ angular distribution (see Sec. \ref{sec:rare_semileptonic_decays}). Again, additional ratio measurements by the LHCb and Belle~II collaborations are eagerly anticipated.

Tests of lepton universality in $\decay{b}{c\ell^-\bar{\nu}_\ell}$ tree-level decays that also show interesting tensions with SM predictions were presented by L.\ Scantlebury-Smead, for details we refer to Ref.~\cite{proceedings:scantleburysmead} and the summary of working group~2 (\vub, \vcb\ and semileptonic/leptonic $B$ decays including $\tau$) at this workshop. 

A theory review of the status of $R_{K^{(*)}}$ uncertainties was given by S.~Nabeebaccus~\cite{proceedings:nabeebaccus}, 
with particular focus on the QED component. 
Although these contributions come with a small coupling $\alpha / \pi \approx 2 \cdot 10^{-3}$, their importance in $\bar B \to \bar K \ell^+ \ell^-$ is due to the presence of kinematic effects enhancing these corrections to ${\cal O}(\alpha / \pi) \log(m_\ell / m_B) \gtrsim 2$-3\%. A first analysis---single-differential in the di-lepton invariant mass squared $q^2$---was performed in Ref.~\cite{Bordone:2016gaq}, finding agreement with PHOTOS~\cite{Davidson:2010ew} at per mil level, and assigning $R_K$ an error of 1\%. Reference~\cite{Isidori:2020acz} presents a more general study of these corrections, with further work in progress. The paper calculates the full matrix elements, \ie\ both the real and virtual components, within an EFT Lagrangian description, i.e scalar QED. The aim is to capture effects beyond collinear $\log(m_\ell / m_B)$ terms. The speaker addresses the question whether the EFT approach employed may miss any $\log(m_\ell / m_B)$ contributions due to structure dependence. Reference~\cite{Isidori:2020acz} argues that this is not the case, on grounds of gauge invariance. On the other hand, this analysis does not capture effects $\propto \log(m_K/m_B)$, which can be non-negligible. A general survey of structure-dependent contributions is on-going within a light-cone sum rules approach.
Work is also in progress as concerns the $\bar B \to \bar K \ell^+ \ell^-$ fully differential distribution from a Monte Carlo. Such a tool can be profitably used to cross-check PHOTOS, and to investigate effects of charmonium resonances.

Open challenges on the crucial topic of QED corrections to $b \to s$ decays were discussed by R.~Szafron~\cite{proceedings:szafron}. 
The speaker first makes two points on the phenomenology side of the problem. First, structure-dependent corrections for semi-leptonic heavy-to-light lepton-universality ratios---including $R_{K^{(*)}}$---depart from unity by terms of $O(\alpha)$ times $O(\log(m_\mu^2 / ...)$, where ellipses denote {\em any} scale in the experimental observable. The worst-case scenario is of an additional few-\% uncertainty from such structure-dependent corrections. The second point is that lattice evaluations of QED corrections for heavy mesons are complementary to the EFT approach. The latter captures terms proportional to $\log(m_\mu / \Delta E) \sim 2.5$ and $\log(m_B / m_\mu) \sim 4$, whereas lattice QCD can estimate $\log(m_B / \Lambda_{QCD}) \sim 3$ terms. Since both classes of terms are of similar size, we should pay attention to both.

\begin{figure}[t]
  \begin{center}
    \includegraphics[width=0.9\textwidth]{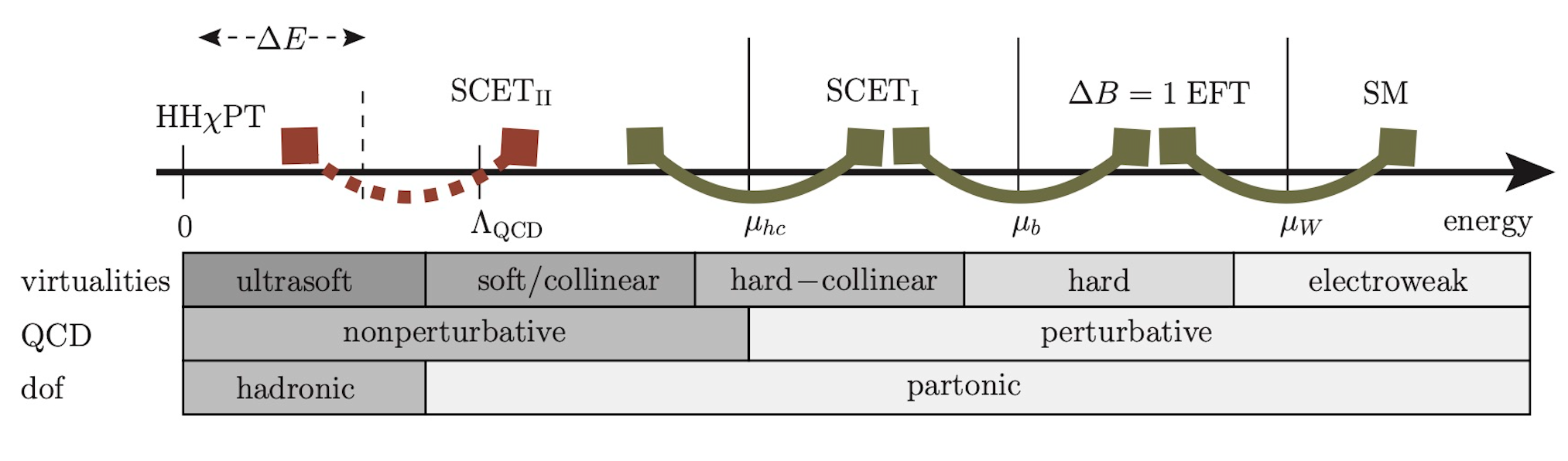}
  \end{center}
  \caption{Tower of theories and scales relevant for the description of semi-leptonic heavy-to-light lepton-universality ratios including QED corrections. $\Delta E$ denotes the minimum energy for single-photon detection. Figure taken from Szafron's contribution to the workshop.
\label{fig:szafron}
}
\end{figure}

On the theory side, the challenge is to perform a non-perturbative matching between the point-like EFT and the microscopic description. This corresponds to a tower of theories that are matched to one another as the energy decreases. The picture is best summarized in Fig.~\ref{fig:szafron}, taken from the seminar. A benchmark application, to $\Bs \to \mu^+ \mu^-$, is Ref.~\cite{Beneke:2017vpq,Beneke:2019slt}, which identifies dangerous single and double $\log(m_b \omega / m_\ell^2)$ terms ($\omega \approx \Lambda_{\rm QCD}$).
After this work, further steps towards a systematic treatment of QED in charmless $B \to \pi^+ \pi^-$ and in heavy-to-heavy decays have been undertaken in Refs.~\cite{Beneke:2020vnb,Beneke:2021jhp}.

An important further challenge concerns non-perturbative soft matrix elements, that need also to be evaluated within QCD$\times$QED, and light-cone distribution amplitudes have to be generalized accordingly. Such generalization has been accomplished in Ref. \cite{Beneke:2021pkl}.

R.~Szafron also emphasizes that dedicated Monte Carlo simulations of QED effects~\cite{Golonka:2005pn} are compatible with the EFT description above $\Lambda_{\rm QCD}$. These Monte Carlos at present neglect radiation from charged initial-state particles and other effects which can be non-negligible. This point, the speaker argues, requires further attention within the community. See also the corresponding discussion in Saad Nabeebaccus' talk. A final, daunting challenge is to go beyond leading power in the $1 / m_B$ expansion.

\subsection{\boldmath Global fits of $\decay{b}{s\ellell}$ data}
\noindent The data on rare $\decay{b}{s\ellell}$ decays is interpreted in an effective field theory framework, whose Wilson coefficients are determined through a global fit, as was discussed in the presentation by P.~Stangl~\cite{proceedings:stangl}. 
The correlation matrix used in these fits includes an experimental and a theory component. Importantly, the latter in general depends on the Wilson coefficients themselves, and as such it may appreciably differ in the SM and in NP scenarios.

A first test is to compare experimental data with observables calculated within the weak effective theory (WET) at the $b$-mass scale of 4.8 GeV. Specifically, one can explore systematically scenarios with shifts to one, or more, Wilson coefficients, with inclusion of theoretically clean observables only, or including also the theoretically less clean ones, in the sense discussed above in connection with $c \bar c$ contributions (see talk by van Dyk~\cite{proceedings:vandyk}). With these different global-fit declinations, one can see that scenarios with shifts to the muonic Wilson coefficients $C_9$, $C_{10}$ and $C_9 = - C_{10}$ have, invariably, pulls well above $3\sigma$. The dependence of the correlation matrix on the Wilson coefficients has been investigated in detail in Ref.~\cite{Altmannshofer:2021qrr} and found to have an impact mostly on the $C_9$-only scenario.

\begin{figure}
    \centering
   \includegraphics[width=0.45\textwidth]{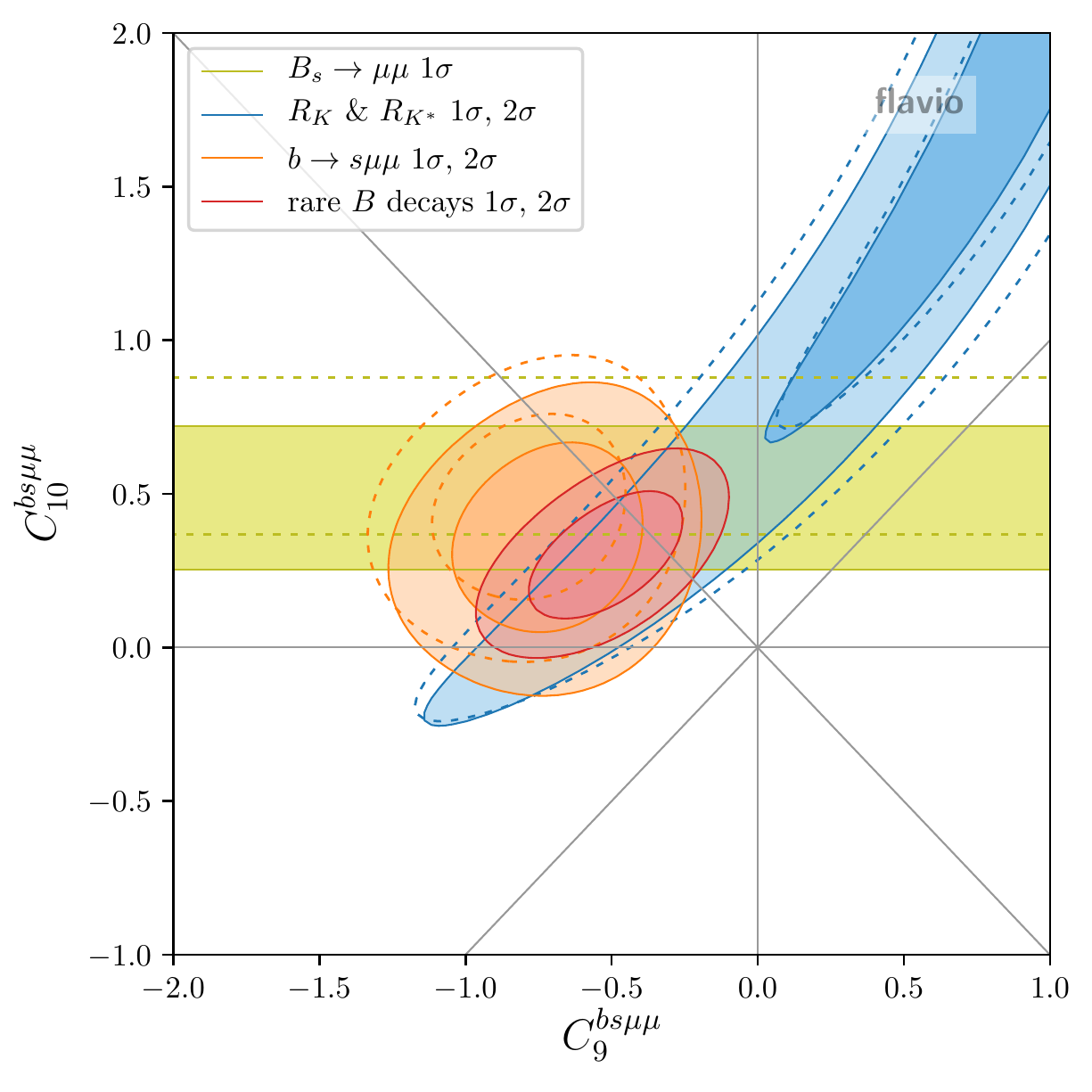} \hfill
   \includegraphics[width=0.49\textwidth]{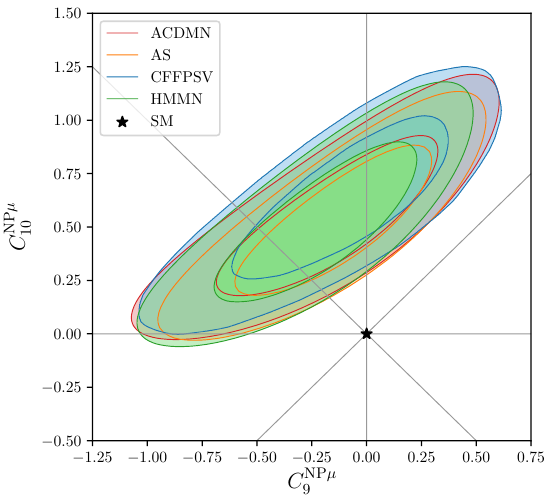}
    \caption{(Left) Fit of the Wilson coefficients $C_9$ and $C_{10}$ using (blue) clean observables and (red) all data on $\decay{b}{s}\ellell$ transitions. Figure from Ref.~\cite{Altmannshofer:2021qrr}. (Right) A comparison of the global-fit regions selected by the approaches in Refs.~\cite{Altmannshofer:2021qrr,Alguero:2021anc,Ciuchini:2020gvn,Hurth:2021nsi} when including only the theoretically cleanest observables, \ie\ LFU ones and $\Bs \to \mumu$. Figure from Ref. \cite{proceedings:stangl}.}
    \label{fig:globalfits}
\end{figure}

Quite intriguing is the picture within global fits to two Wilson coefficients, in particular in the plane $C_9$ vs. $C_{10}$, see Fig.~\ref{fig:globalfits}. Here one may compare a few reference hypotheses on the data: a first one consists in including only $\Bs \to \mumu$ as well as the LFU observables $R_K$, $R_{K^*}$ and $D_{P_{4,5}}^\prime$~\cite{Altmannshofer:2015mqa,Capdevila:2016ivx,Serra:2016ivr}. This case prefers muonic-only Wilson coefficients. In a second reference scenario one considers $b \to s \mumu$ data---\ie\ branching fractions and angular observables---with the caveat that they may be affected by possibly underestimated hadronic uncertainties. Agreement between the datasets in these two scenarios is improved by a flavour-universal shift to $C_9$ \cite{Alguero:2018nvb}, usually denoted as $C_{9}^{\rm univ}$. Interestingly, such shift lends itself to a natural interpretation~\cite{Bobeth:2011st,Crivellin:2018yvo} that can quantitatively connect $b \to s \ellell$ and $b \to c \ell^- \bar{\nu}$ anomalies~\cite{Aebischer:2019mlg}. Combining the datasets from the two scenarios (global fit) yields a best-fit region perfectly compatible with the line $C_9 = - C_{10}$, with a negative central value.

Another interesting highlight is that global-fit conclusions are robust across collaborations. Quite impressive in this respect is an explicit comparison between the approaches in Refs.~\cite{Altmannshofer:2021qrr,Alguero:2021anc,Ciuchini:2020gvn,Hurth:2021nsi} in the mentioned $C_{9}$ vs. $C_{10}$ plane. Within a fully global fit, the different approaches yield partly overlapping regions, whose location can be understood in light of the somewhat different treatment of certain observables across the approaches. For example, a more or less conservative stance on angular data in $B \to K^* \ellell$ leads to a less or more pronounced preference for the $C_9$-only scenario with respect to the competitor $C_9 = - C_{10}$. However, and crucially, if one focuses attention on theoretically clean observables only, to wit LFU ones and $\Bs \to \mumu$, all approaches lead to, basically, the very same region in the $C_9$ vs. $C_{10}$ plane, as shown in Fig. \ref{fig:globalfits} (right). This region prefers a non-zero shift to one or both of these Wilson coefficients, with the SM point about 4$\sigma$ away from the best-fit point.

\subsection{Model building for flavour anomalies} \label{sec:model-building}
\noindent Model building for the flavour anomalies was discussed by C.~Cornella~\cite{proceedings:cornella}. 
She first notes that $b \to s$ anomalies point to a scale anywhere between a few and about 40 TeV, whereas $b \to c$ discrepancies, if confirmed, require a lower scale, around a few TeV. An interesting basic question is why do we want to consider both sets of discrepancies on the same footing. The immediate answer is that this is natural because the two underlying quark currents are related by $SU(2)_L$ gauge invariance, which points to left-handed semileptonic operators. (Although extensions with right-handed currents are also possible.) These features are suggestive of UV completions becoming dynamical somewhere above the SMEFT scale, where the new dynamical scale is integrated out, but the full SM gauge symmetry is still in place.

Cornella outlines the three possibilities that stand out for a combined explanation: the two scalar leptoquarks known as $S_1$ and $S_3$ (see in particular Refs.~\cite{Crivellin:2017zlb,Buttazzo:2017ixm,Marzocca:2018wcf}); the alternative scalar-leptoquark combination of $R_2$ and $S_2$~\cite{Becirevic:2018afm}; the vector leptoquark $U_1$ (see in particular Refs.~\cite{DiLuzio:2017vat,Calibbi:2017qbu,Bordone:2017bld,Barbieri:2017tuq,Heeck:2018ntp}). In this case, there are automatically no tree-level contributions to $b \to s \nu \bar \nu$. The $U_1$ scenario is certainly among the favorite ones (for a recent comparison of the description of current data across the different scenarios, see Ref.~\cite{Angelescu:2021lln}) and is characterized by well-defined low-energy predictions, including a large modification of $b \to s \tau^+ \tau^-$ currents, as well as large $\tau / \mu$ lepton flavour violation in $b \to s \tau \mu$ currents as well as $\tau$ decays~\cite{Cornella:2021sby}. Importantly, this scenario can also be probed at high $p_T$ through di-tau tails~\cite{Faroughy:2016osc,Cornella:2021sby}, and if this scenario should be part of the answer, excesses in $pp\to\tau^+ \tau^-$ are to be expected sooner or later.

Another important point made is that the $U_1$ case has inspired numerous attempts towards a full-fledged, renormalizable model. This has happened to some extent by necessity: there exist observables that feature a power-like sensitivity to the UV scale and crave for a UV completion---for a neat early discussion see Ref.~\cite{Barbieri:2015yvd}, whereas the most up-to-date treatment is to be found in Ref.~\cite{Fuentes-Martin:2020hvc}. In short, efforts towards a UV completion of the $U_1$ simplified model are especially important in order to rigorously test this framework. There are different possible paths towards this end, and here we limit ourselves to  a short summary of the gauge path. The starting point is, not surprisingly, the Pati-Salam model~\cite{Pati:1974yy}---after all, a leptoquark vector mediator immediately calls for the idea of lepton number as the fourth color. The original Pati-Salam group, however, faces a number of challenges in light of the wealth of existing collider data, as discussed in Ref.~\cite{DiLuzio:2017vat}. In particular, one has to disentangle the $SU(4)$ group from $SU(3)_c$, and one has to circumvent high-$p_T$ constraints by requiring $g_4 \gg g_1$, $g_3$. The ensuing construction is the so-called 4321 model~\cite{Georgi:2016xhm,Bansal:2018eha,DiLuzio:2017vat}. An important aspect of this model is that the $U_1$ does not come alone as a mediator. It is accompanied by a $Z^\prime$ and a ``coloron'' field, whose phenomenology is also to be taken into account~\cite{Cornella:2021sby}. Finally, this short summary has to skip other more recent model-building directions yielding leptoquarks, for which the reader is referred to Ref.~\cite{proceedings:cornella}.

\section{\boldmath Rare $D$ decays}
\noindent Rare and forbidden charmed meson decays, such as FCNC $|\Delta C| = 1$ or purely leptonic transitions, are unique probes for physics beyond the SM in the up-type quark sector and are complementary to analogous searches in the $K$ and $B$ sector.

Recent results on rare and forbidden semileptonic $D_{(s)}$ decays at BES~III were discussed by L.~Sun~\cite{proceedings:sun}. 
The search for the semileptonic decay $D_s \to \gamma e \nu_e$ is reported in Ref.~\cite{BESIII:2019pjk}. No evidence for a signal was found and an upper limit on the branching fraction of $1.3 \times 10^{-4}$ at $90\%$ CL was set. This represents the first search for this mode and has to be compared with a SM prediction of $3 \times 10^{-5}$. An indirect search for Majorana neutrinos in $D \to K \pi e^+e^+$ decays was also presented~\cite{BESIII:2019oef}. Upper limits on the branching fraction for the SM process and as a function of the Majorana neutrino mass were set at 90\% confidence level. The former varies from $2.8 \times 10^{-6}$ for the $D^0 \to K^- \pi^- e^+e^+$ mode to $8.5 \times 10^{-6}$ for the decay $D^+ \to K^- \pi^0 e^+e^+$ while the latter ranges between $10^{-7}$ and $10^{-2}$ depending on the new state mass and on the reconstructed final states.
Lastly, upper limits of the order of $10^{-6}$ to $10^{-7}$ on a search for baryon number violation with $D \to \Lambda(\Sigma_0) e$ were presented~\cite{BESIII:2019udi}.
The $20\invfb$ dataset collected in the next few years by BES~III at the $\Psi(3770)$ threshold will allow to carry out a rich program on rare and forbidden charm decay searches.

 LHCb searches on rare and forbidden $D$ decays were covered by D.~Brundu~\cite{proceedings:brundu}. Among the many accessible modes, the following ones were discussed: $D_{(s)}^+ \to h^+ \ell^\pm \ell^{(\prime)\mp}$~\cite{LHCb:2020car} and $D^0 \to h^+h^- \mumu$~\cite{LHCb:2021yxk}, with $h$ being a pion or a kaon. 
 In the search of the 25  $D_{(s)}^+ \to h^+ \ell^\pm \ell^{(\prime)\mp}$ modes, no evidence for signal was found and upper limits on the branching fractions were set, with an improvement of more than one order of magnitude, apart from the $h e^+ e^-$ final states, with respect to previous measurements.
The $D^0 \to h^+h^- \mumu$ decays were studied to measure the CP asymmetry and angular variables as a function of the di-muon invariant mass. 
All measured quantities have been found to be in agreement with SM predictions and, since they are limited by the size of the data samples, will benefit from updated analyses using more data. 

\section{\boldmath Rare $K$ decays}
\noindent Rare $K$ decays are well-known to rank among the most sensitive flavour constraints of physics beyond the SM. Precision measurements and new physics searches in the kaon sector were discussed, both from the theoretical and the experimental point of view. Here we shortly review the two aspects in turn.

M.~Bruno~\cite{proceedings:bruno} discussed the lattice-QCD (LQCD) status of $\epsilon^\prime / \epsilon$, $\epsilon_K$, as well as $K \to \pi \nu \bar \nu$. This is a vast topic, where major progress has been made over recent years in LQCD calculations~\cite{Aoki:2021kgd}.
As regards $\epsilon^\prime / \epsilon$, it is well-known that the crucial challenge is to compute the imaginary parts of the building-block amplitudes $A_{0,2} \equiv A(K \to (\pi \pi)_{I=0,2})$, with the $\pi \pi$ in a state of definite isospin $I$. One further challenge used to be the calculation of the corresponding strong-phase difference $\delta_2 - \delta_0$, performed with either dispersive~\cite{Colangelo:2001df} or lattice~\cite{RBC:2020kdj} approaches. These two approaches are now in perfect agreement with each other. In synthesis, one can obtain an estimate of Re($\epsilon^\prime / \epsilon$) by taking  Re$A_0$, Re$A_2$, as well as $\omega =$Re$A_2/$Re$A_0$ (the ``$\Delta I = 1/2$ rule'') from experiment~\cite{ParticleDataGroup:2020ssz}; the $\delta_2 - \delta_0$ phase difference can be taken from either a dispersive or a lattice approach as mentioned; finally, one can take Im$A_2$ from Ref.~\cite{RBC:2015gro} and Im$A_0$ from the more recent calculation in Ref. \cite{RBC:2020kdj}. The ensuing prediction reads \cite{RBC:2020kdj}
\be
\label{eq:e'/e}
{\rm Re}(\epsilon^\prime / \epsilon) = 21.7(2.6)(6.2)(5.0) \times 10^{-4}~,
\ee
to be compared with the experimental result of $16.6(2.3) \times 10^{-4}$ \cite{ParticleDataGroup:2020ssz}. In eq.~(\ref{eq:e'/e}) errors are statistical, systematic from sources other than the isospin-breaking and e.m.\ corrections, and systematic due to these two components. It is interesting to note that, if one applied to this result the (negative) correction from isospin breaking calculated in Ref.~\cite{Cirigliano:2019cpi}, the central value in eq.~(\ref{eq:e'/e}) would basically coincide with the experimental result.

LQCD also produced first numerical results at $m_\pi^{\rm phys}$ for $\Delta m_K$ \cite{Wang:2019try} (see also Refs.~\cite{Christ:2012se,Bai:2014cva,Christ:2015pwa}), with the preliminary result $\Delta m_K = 6.7(1.7) \times 10^{-12}$ MeV, to be compared with the experimental measurement of $3.483(6) \cdot 10^{-12}$ MeV. The preliminary lattice result has a somewhat intriguing central value, but also a  large uncertainty, dominated by discretization errors from charm. It is worth noting that the $m_\pi^{\rm phys}$ limit requires large spatial sizes and inclusion of charm, which in turn demands fine lattice spacing $a$---a challenging task.

In $\epsilon_K$ the current frontier in LQCD is the calculation of long-distance contributions, subsumed in the parameter $\xi$. The most recent result reads $\xi = 0.17(1) \cdot 10^{-3}$~\cite{Bai:2016gzv}. The underlying calculation is challenging for many reasons, in particular the large number of contractions and topologies. This result may be compared with the phenomenological estimate in Ref.~\cite{Buras:2008nn}---which relies on the relationship between $\xi$ and the experimental value of Re($\epsilon^\prime / \epsilon$) and hence assumes the absence of new-physics contributions on the latter---as well as with the ChPT calculation in Ref.~\cite{Buras:2010pza}, which includes only the dominant non-analytic terms---and assumes nothing on $\epsilon^\prime / \epsilon$. A lattice approach to $\xi$ is an important milestone.

One further LQCD highlight at CKM2021 was the status of $K \to \pi \nu \bar \nu$. This decay is dominated by short-distance effects. In fact, starting from the relevant Hamiltonian~\cite{Buchalla:1993wq,Buchalla:1998ba}, the (non-perturbative) hadronic matrix element is related by isospin \cite{Mescia:2007kn} to the experimentally measured  $K_{\ell3}$ decay. However, long-distance effects may be up to 6\% in the charged mode $K^+ \to \pi^+ \nu \bar \nu$ \cite{Isidori:2005xm}. Exploratory studies \cite{Bai:2017fkh,Bai:2018hqu} at unphysical kinematics observe a curious cancellation between the contributions from each of the $WW$ and $Z$-exchange diagrams to the difference between the full LQCD result for the charm-quark contribution to the amplitude and the amplitude obtained using perturbation theory. A  more recent calculation \cite{Christ:2019dxu} was carried out at the near-physical pion mass of $m_\pi = 170$ MeV (and unphysical charm), which shows a mild momentum dependence (varying momenta in LQCD is computation-intensive) and clarifies the role of intermediate $\pi \pi$ states. These are technical steps towards the ultimate aim---to calculate the full rate with below-percent precision at physical kinematics.

A. Buras ~\cite{Buras:2022cyc} discusses the theory status, both within and beyond the SM, of the $\epsilon^\prime / \epsilon$ anomaly, which was the object of a controversy till 2020. The experimental average from NA48 and KTeV reads Re$(\epsilon^\prime / \epsilon) = 16.6(2.3) \cdot 10^{-4}$. \cite{NA48:2002tmj,KTeV:2002qqy,Worcester:2009qt}. A first theory comparison can be made with ChPT, whose most recent determination yields Re$(\epsilon^\prime / \epsilon) = 14(5) \cdot 10^{-4}$ \cite{Cirigliano:2019ani,Gisbert:2020wkb} and showing perfect agreement. As discussed earlier, LQCD has also accomplished a full calculation of the same quantity, and the latest result in eq. (\ref{eq:e'/e}) shows, again, good agreement with the experimental measurement. Finally, a fully analytic calculation \cite{Buras:2015xba,Buras:2016fys} can also be performed within the context of so-called ``dual QCD'' \cite{Bardeen:1986vz,Buras:2014maa}, which is grounded on a large-$N_c$ expansion of QCD. The prediction reads Re$(\epsilon^\prime / \epsilon) = 5(2) \times 10^{-4}$ \cite{Buras:2020wyv} which at face value signals a tension with respect to the experimental result.

The talk also summarizes the outstanding tasks ahead for each of the three mentioned theory approaches. On the LQCD side, and taking as reference the RBC-UKQCD setup, the next milestones are the calculation of isospin-breaking as well as QED corrections---that produce the last error component in eq.~(\ref{eq:e'/e}). A second, significant source of systematics is the finite lattice spacing. See Ref.~\cite{RBC:2020kdj} for a detailed discussion. On the ChPT side, the current uncertainty is dominated by the input values of the strong low-energy constants, in particular $L_{5,7,8}$ and by the poor knowledge of $1/N_c$-suppressed contributions in the matching region---at present estimated conservatively by scale variations \cite{Gisbert:2020wkb}. Finally, within the DQCD calculation, important work is necessary in order to include subleading final-state interactions.

D. Marzocca~\cite{proceedings:marzocca} provided a theory perspective on the possibility of correlations between $K \to \pi \nu \bar \nu$ and the $B$ anomalies. First insights can be obtained already at the EFT level, with general, symmetry-guided assumptions on the flavour structure. Such structure can then be fixed completely within specific models.

As regards the EFT level, one should first note that $B$ anomalies suggest new physics dominantly coupled to the third generation of down-type fermions \cite{Glashow:2014iga}. This, by way of flavour mixing after electroweak-symmetry breaking, implies dominant flavour effects in $b \to s$ transitions (and in final states with $\tau$, including lepton-flavour violating ones); however, because of this very mixing, effects in general percolate to any other flavour combination.\footnote{This observation was made properly $SU(2)_L$-symmetry compliant in Ref. \cite{Bhattacharya:2014wla} thus paving the way for joint explanations of $b \to s$ and $b \to c$ data alike (see also introductory discussion in Ref. \cite{Greljo:2015mma}).} This simple picture encounters well-defined limitations in the light of data, as discussed in Ref.~\cite{Buttazzo:2017ixm}, which also offered a number of paths forward. A well-defined direction is that of considering a minimally-broken $U(2)^5$ global symmetry \cite{Barbieri:2011ci,Barbieri:2012uh}. Such assumption offers a natural interpretation of the resemblance between 
the seemingly hierarchical nature of NP couplings to the three generations and the hierarchy in SM fermion masses. Starting from this ansatz, one can relate, even within an EFT context, flavour effects across different generations up to $O(1)$ coefficients. Such approach was utilised to study the possible correlation between $R(D^{(*)})$ and BR$(K^+ \to \pi^+ \nu \bar \nu)$ as a function of the undetermined $O(1)$ coefficients \cite{Bordone:2017lsy} (see also \cite{Fajfer:2018bfj}). The two processes suggest new scales of, respectively, $O(4~\mbox{TeV})$ and 42 TeV. It is interesting that the latter scale is very close to the general EFT bound implied by the NA62 measurement of BR$(K^+ \to \pi^+ \nu \bar \nu)$ \cite{NA62:2021zjw}. In a similar vein, Ref. \cite{Borsato:2018tcz} related semi-leptonic $B$ anomalies with lepton-flavour violating $K$ decays of the kind $K \to (\pi) e^\pm \mu^\mp$.

More accurate correlations between different flavour sectors can be obtained by committing to full-fledged models, with new particle d.o.f. such as leptoquarks (LQs). As discussed in Cornella's talk at CKM2021, three main LQ sets stand out for combined explanations of the $B$ anomalies, see Sec. \pageref{sec:model-building} form more details. The cases of scalar LQs have the advantage of being fully calculable already at the simplified-model level. This is not the case for vector LQs, as also discussed there. Other interesting features of scalar-LQ scenarios is that they can address $(g-2)_\mu$; finally, their UV origin may potentially be found in composite-Higgs models, in turn interesting for a possible connection with the EW hierarchy problem \cite{Marzocca:2018wcf}.

Focusing on the instance of the $S_1 + S_3$ model, one may perform a detailed analysis of correlations between $B$- and kaon-physics observables. After matching the SM plus the $S_1 + S_3$ simplified model to SMEFT at 1 loop \cite{Gherardi:2020det}, one can first perform a joint analysis of $B$ anomalies, thereby fixing the relevant couplings \cite{Gherardi:2020qhc}, and subsequently include 1$^{\rm st}$-generation couplings, in order to study kaon and $\mu \to e$ observables. This inclusion requires assumptions on the underlying flavour structure, e.g. a flavour symmetry. Reference~\cite{Marzocca:2021miv} considers in detail the case of a $U(2)^5$ flavour symmetry, already mentioned above. Then, e.g. the left-handed LQ couplinga to the down quark are equal to the strange-quark counterparts up to the small CKM factor $V_{\rm td} / V_{\rm ts}$. This establishes a strong correlation between $K$ and $B$ decays. A global study finds that the $S1 + S3$ LQs can accommodate $R(D^{(*)})$ only at $2\sigma$, because of the stringent constraints from BR$(K^+ \to \pi^+ \nu \bar \nu)$, $\epsilon_K$ and $Z \to \tau \bar \tau$.

An experimental overview on $K \to \pi \nu \bar\nu$ searches was presented by Y.-C.~Tung~\cite{proceedings:tung}. 
The most precise measurement on the charged channel has been performed by the NA62 collaboration~\cite{NA62:2021zjw} using the 2017--2018 dataset. An excess of events with respect to the background-only prediction at the 3.4$\,\sigma$ level was observed. The corresponding measured branching fraction, at 68\% confidence level, is:
\begin{equation}
    \mathcal{B}(K^+ \to \pi^+ \nu \bar\nu) = (10.6 ^{+4.0}_{-3.4}(\mbox{stat})\pm 0.9 (\mbox{syst})) \times 10^{-11},
\end{equation}
consistent with the SM expectation within errors. 
NA62 has resumed data-taking in mid 2021, and, with the sample collected by end 2024, will be able to push down the statistical precision to the 10\% level
thanks to improvements in the detector, beam line and beam intensity.

The neutral channel has not been observed yet, the current best limit at 90\% confidence level has been set by the KOTO collaboration with an analysis based on data collected in 2015~\cite{KOTO:2018dsc}:
\begin{equation}
    \mathcal{B}(K^0_L \to \pi^0 \nu \bar\nu) < 3.0 \times 10^{-9}~,
\end{equation}
to be compared with the SM prediction $(3.00 \pm 0.30) \times 10^{-11}$~\cite{Buras:2015qea}. 
Data from 2016--2018 runs have also been analyzed~\cite{KOTO:2020prk}, yielding three events in the signal region. Detailed studies on additional background sources were performed after the signal region opening and two sources not previously accounted for were found: $K^0_L \to \gamma\gamma$ from beam halo, with two photons faking the neutral pion,  and $K^+ \to e \gamma\nu$, with the electron escaping the calorimeter acceptance. The above result is found to be consistent with the estimated total number of background events, $1.22 \pm 0.26$.  
An upgrade at both J-PARC beam facility and KOTO are taking place, data taking will resume in 2022 and by 2026 a sample 11 times larger will be collected. 
A major upgrade from KOTO to KOTO STEP-II~\cite{Aoki:2021cqa} is foreseen after 2026 with start of physics runs foreseen for 2029.

A review of recent searches performed at NA62 in modes other than $K^+ \to \pi^+ \nu \bar{\nu}$ was given by E.~Goudzovski~\cite{proceedings:goudzovski}.
The $K^+ \to \pi^+ \nu \bar\nu$ and $K^+ \to \pi^+ \pi^0$ analyses were extended to search for $K^+ \to \pi^+ X$~\cite{NA62:2020xlg} and $\pi^0 \to X$~\cite{NA62:2020pwi}, with $X$ being a new particle decaying to invisible final states.
Searches for heavy neutral leptons $N$ have been performed reconstructing
$K \to \ell N$ final states~\cite{NA62:2020mcv}~\cite{NA62:2021bji} resulting in improvements  in limits on the couplings of up to two orders of magnitude with respect to previous works.
This allows to search in different $X$ mass regions with limits on the $X$ couplings exceeding previous works in the sub-100 $\mevcc$ range.
Six lepton number and lepton flavor violating modes were searched for, 
including $K \to \pi \ell^+\ell^+$ and $K \to \pi \ell^+ \ell^{'-}$, 
all producing limits on the branching fractions as strong as $10^{-11}$~\cite{NA62:2019eax,NA62:2021zxl}. Updated preliminary results on event yield for the rare decays $K \to \pi \mu^+ \mu^-$ and  
$K \to \pi e \nu \gamma$ were also reported.

\section{Acknowledgments}

\noindent We would like to thank all the speakers for contributing extremely interesting talks, and Phillip Urquijo for the non-trivial organization and logistics. The speaker would like to thank Elisa and Christoph for a very enjoyable collaboration.
Christoph and Elisa would like to thank Diego as well for the enjoyable collaboration and for brilliantly delivering the WG summary.

%

\bibliographystyle{JHEP}
\bibliography{bibliography}

\end{document}